\documentclass[sigconf]{acmart}

\copyrightyear{2024}
\acmYear{2024}
\setcopyright{acmlicensed}
\acmConference[UIST '24]{The 37th Annual ACM Symposium on User Interface Software and Technology}{October 13--16, 2024}{Pittsburgh, PA, USA}
\acmBooktitle{The 37th Annual ACM Symposium on User Interface Software and Technology (UIST '24), October 13--16, 2024, Pittsburgh, PA, USA}
\acmDOI{10.1145/3654777.3676391}
\acmISBN{979-8-4007-0628-8/24/10}

\usepackage[ruled]{algorithm2e}
\usepackage{listings}
\usepackage{xcolor}

\colorlet{punct}{red!60!black}
\definecolor{background}{HTML}{f6f6f6}
\definecolor{delim}{RGB}{20,105,176}
\colorlet{numb}{magenta!60!black}

\lstdefinelanguage{json}{
    basicstyle=\normalfont\ttfamily,
    numberstyle=\scriptsize,
    showstringspaces=false,
    breaklines=true,
    frame=lines,
    backgroundcolor=\color{background},
    literate=
     *{0}{{{\color{numb}0}}}{1}
      {1}{{{\color{numb}1}}}{1}
      {2}{{{\color{numb}2}}}{1}
      {3}{{{\color{numb}3}}}{1}
      {4}{{{\color{numb}4}}}{1}
      {5}{{{\color{numb}5}}}{1}
      {6}{{{\color{numb}6}}}{1}
      {7}{{{\color{numb}7}}}{1}
      {8}{{{\color{numb}8}}}{1}
      {9}{{{\color{numb}9}}}{1}
      {:}{{{\color{punct}{:}}}}{1}
      {,}{{{\color{punct}{,}}}}{1}
      {\{}{{{\color{delim}{\{}}}}{1}
      {\}}{{{\color{delim}{\}}}}}{1}
      {[}{{{\color{delim}{[}}}}{1}
      {]}{{{\color{delim}{]}}}}{1},
}

\begin{document}

\newcommand{\sys}{ProgramAlly\xspace}

\newcommand{\add}[1]{#1}


\title{\sys: Creating Custom Visual Access Programs via Multi-Modal End-User Programming}


\author{Jaylin Herskovitz}
\affiliation{%
 \institution{University of Michigan}
 \city{Ann Arbor, MI}
 \country{USA}}
\email{jayhersk@umich.edu}

\author{Andi Xu}
\affiliation{%
 \institution{University of Michigan}
 \city{Ann Arbor, MI}
 \country{USA}}
\email{andixu@umich.edu}

\author{Rahaf Alharbi}
\affiliation{%
 \institution{University of Michigan}
 \city{Ann Arbor, MI}
 \country{USA}}
\email{rmalharb@umich.edu}

\author{Anhong Guo}
\affiliation{%
 \institution{University of Michigan}
 \city{Ann Arbor, MI}
 \country{USA}}
\email{anhong@umich.edu}


\begin{abstract}

Existing visual assistive technologies are built for simple and common use cases, and have few avenues for blind people to customize their functionalities. Drawing from prior work on DIY assistive technology, this paper investigates end-user programming as a means for users to create and customize visual access programs to meet their unique needs. We introduce \sys, a system for creating custom filters for visual information, e.g., \texttt{`find NUMBER on BUS'}, leveraging three end-user programming approaches: block programming, natural language, and programming by example. To implement \sys, we designed a representation of visual filtering tasks based on scenarios encountered by blind people, and integrated a set of on-device and cloud models for generating and running these programs. In user studies with 12 blind adults, we found that participants preferred different programming modalities depending on the task, and envisioned using visual access programs to address unique accessibility challenges that are otherwise difficult with existing applications. Through \sys, we present an exploration of how blind end-users can create visual access programs to customize and control their experiences.
\end{abstract}




\begin{teaserfigure}
\centering
\vspace{-0.5pc}
  \includegraphics[width=\textwidth]{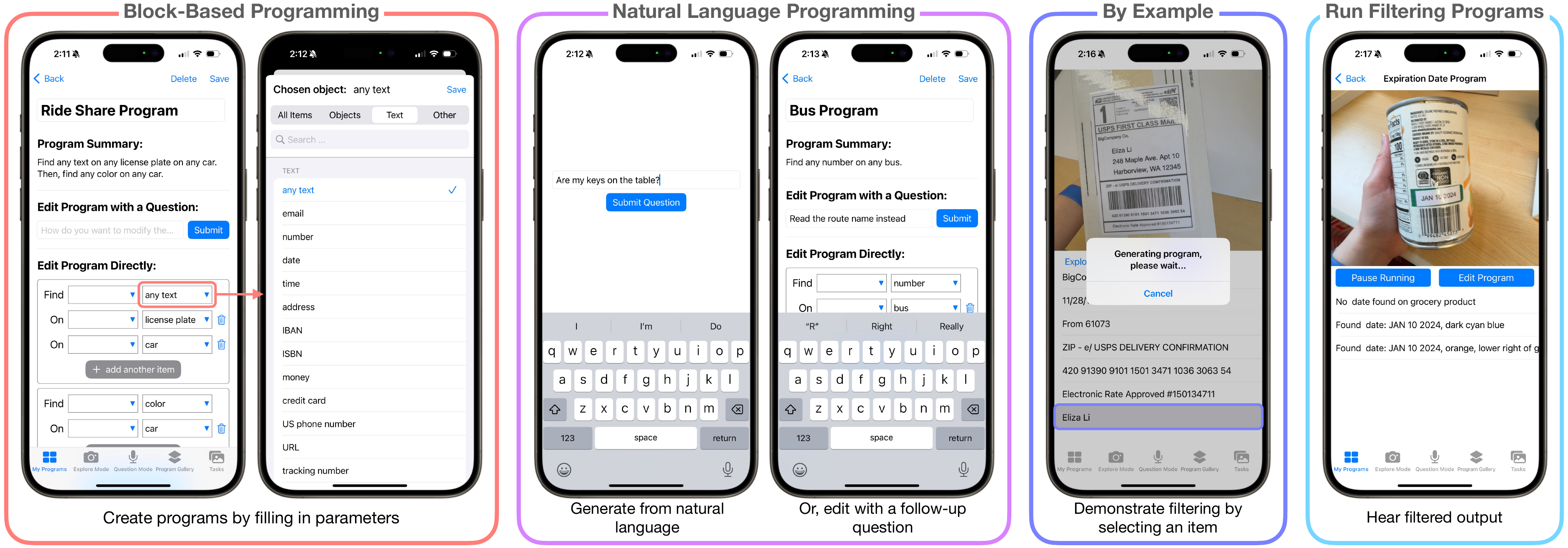}
  \vspace{-1pc}
  \caption{\sys is an end-user programming tool for creating visual information filtering programs. \sys provides a multi-modal interface, with block-based, natural language, and programming by example approaches. 
  }
  \vspace{0.5pc}
  \Description{Figure 1: An overview of ProgramAlly's interface. (1) Block-Based programming mode. This shows a program called 'ride share program'. The screen has a program summary which reads: 'Find any text on any license plate on any car. Then, find any color on any car'. Below that there is a text box which says: 'Edit program with a question'. Finally, the screen shows each 'Find' and 'On' statement in the program and the interface for editing them. For VoiceOver users, this sounds like: 'Find any text, actions available. Edit adjective. Edit object text. Delete item'. (2) The next screen shows the menu that appears when activating one of the edit actions. It has a list of items like 'any text', 'email', 'number', etƒc. (3) Natural language programming mode. It shows a text box that reads 'Are my keys on the table?' with a button that says 'submit question'. (4) Programming by example mode. The interface shows a camera view on top, displaying a package. Underneath is a list of all text and items found in the camera feed that shows 'Eliza Li' being selected. Overlaid on the screen is an alert that says 'Generating program, please wait'. Underneath, the caption says 'demonstrate filtering by selecting an item'. (5) Shows the app running a filtering program. The title reads 'Expiration date program'. In the camera feed, there is a can of beans with an expiration date on the label. The program output reads 'Found date, JAN 10 2024'.}
  \label{fig:teaser}
\end{teaserfigure}

\maketitle
\vspace{-0.25pc}
\section{Introduction}

Artificial intelligence (AI)-based assistive technologies can help blind people gain visual access in a variety of common scenarios, such as reading printed text and identifying objects. These applications tend to be designed for simple and common use cases to maximize their broad usability, and prior work has demonstrated that there is still a long-tail of diverse scenarios that automated assistive technologies cannot account for \cite{herskovitz2023hacking}.
This leads to users having to shoulder additional cognitive load and adjust how they use the technology to get usable results.
Depending on the application, users may need to sift through irrelevant or repetitive information, ask follow up questions, or re-take photos to find specific pieces of information they are looking for.
While these methods can be acceptable in some situations, they can be especially difficult in situations where people want a specific piece of information quickly.
For example, situations that are repetitive (like sorting mail), time-sensitive (like catching a bus), or require scanning an entire object or room with the camera (like finding an expiration date) can all become burdensome with general purpose assistive technology. 

Do-It-Yourself (DIY) assistive technology research has sought to address the related issue of a lack of customizability in assistive devices \cite{hurst2011empowering}. 
To this end, a variety of approaches have been developed aiming to make it easier for non-experts to create adaptive 3D models for themselves or a family member \cite{hurst2013making, chen2016reprise}. 
The same concept has yet to be fully applied to the space of assistive software.
From prior work, we know that blind people already put a significant degree of effort into customizing, hacking, or simply envisioning new assistive technologies \cite{herskovitz2023hacking}.
Yet, there is a gap between the technologies and customizations that people desire to create, and the systems that can support them in doing so with various degrees of technical expertise. 

End-user programming is a potential method for supporting users in customizing and DIY-ing AI assistive software.
Ko et al. define end-user programming as a form of programming done by non-professionals, `to support some goal in their own domains of expertise', further, `to achieve the result of a program primarily for personal, rather [than] public use' \cite{ko2011state}. This definition is aligned with assistive technology needs: blind people are domain experts in designing and using assistive technology \cite{bigham2018learning}, and there is a long-tail of unique scenarios that require personalization to meet individual needs. 
End-user programming approaches, while powerful for enabling users to do more complex tasks, have not yet been applied to the domain of visual accessibility.
Doing so presents new research challenges, namely in making tools that are approachable, accessible, and expressive.
In this work, we demonstrate the potential of end-user programming approaches for assistive technology creation and customization.

We introduce \sys, an end-user programming tool for creating and customizing reusable visual information filters (see Figure \ref{fig:teaser}).
\sys is a mobile application that provides a multi-modal interface for creating and iteratively editing short block-based programs (e.g., \texttt{`find NUMBER on BUS'}). It is built on a generalizable program representation of similar filtering tasks, derived from a dataset of real-world scenarios from blind people's everyday experiences.
\sys provides a set of methods for implementing programs, using multiple interaction modes: direct input, speech, and camera input. These modes implement common end-user programming approaches: block-based programming, natural language programming, and programming by example.
\sys integrates a set of on-device and cloud models for generating and running programs, and can easily be extended to support new, specialized models.

In a study of \sys with 12 blind participants, we assessed its three program creation modes, comparing \sys to existing AI-powered assistive applications, and gathering participants' thoughts on programming and DIY-ing assistive technology more broadly.
Four of these participants were consulted as \sys was being developed, providing design feedback and suggestions for new features, as well as evaluating the programming interfaces and concept.
The remaining participants performed a final evaluation of \sys in both in-person and remote settings.

We found that participants were receptive to the idea of customizing and programming their assistive technology, even if they had no programming experience.
Participants envisioned using different programming interfaces depending on the program they wanted to write, the setting, and their experiences with technology.
We observed that each interface requires different cognitive and technical skills, and outline specific challenges faced by blind end-user programmers when creating visual programs.

Overall, \sys is an investigation of how end-user programming techniques can be used to create and customize AI-based assistive technology. \sys aims to inform how AI models may be directly used as building blocks by blind people in order to support new, complex tasks. This work aims to promote the democratization of AI technology creation and support blind people in having greater control over the AI-based technologies in their lives.
This paper makes the following contributions:
\begin{enumerate}
 \item A generalized representation of visual information filtering tasks, informed by real-world scenarios from blind people's everyday experiences, that can be easily extended to support new object classes.
 \item \sys, a system instantiating this representation and providing a set of multi-modal interaction methods for creating visual information filtering programs: block-based programming, natural language programming, and programming by example.
 \item A study of \sys with blind users, assessing the application of end-user programming approaches to the DIY assistive technology space and highlighting new challenges faced by blind end-user programmers.
\end{enumerate}

\section{Related Work}

\sys builds upon a body of prior research on accessibility and programming tools.
We first review the need to express specific intents in assistive technology. Then, we review various approaches to technology personalization: personalization in assistive technologies, DIY assistive technology, and end-user programming.

\subsection{Information Seeking in Assistive Technology}

Searching visual scenes for specific pieces of information has always been an important aspect of assistive technology design.
In early remote human assistance approaches like VizWiz, users would submit a question along with an image, and assistants would use their human intelligence to determine a relevant answer \cite{bigham2010vizwiz}.
This need for specific information is present across a variety of accessibility contexts: Find My Things and Kacorri use teachable object recognizers to help blind users locate specific possessions \cite{findmythings, kacorri2017teachable}, VizLens helps blind users search for specific buttons on physical interfaces \cite{guoVizLens2016}, and CueSee highlights products of interest for people with low vision \cite{zhao2016cuesee}.
Even outside of accessibility contexts, the Ctrl-F shortcut for `find' is ubiquitous.
General-purpose and specific assistive technologies each have their uses; compare the ambient audio cues in Microsoft's Soundscape \cite{microsoft-soundscape} to navigation directions from Google Maps, neither is a direct replacement for the other.

Yet, current automated assistive applications present challenges to getting specific information quickly.
Whether they run on a live camera feed (e.g, Seeing AI \cite{microsoft-seeingAI}) or on a static image (e.g, the GPT-4 powered `Be My AI' \cite{bemyai}), commercial applications have taken a general approach to describing visual information, conveying all results from the underlying OCR or object detection models, or generating as rich of a description about the visual content as possible.
While this is sometimes desirable, it risks slowing down and increasing the cognitive burden on users who are looking for something specific \cite{gamage2023blind}.
In this work, we aim to target this need for specificity. \sys is a live assistive technology that can provide continuous feedback, but it also aims to capture a user's explicit intent through the creation of filtering programs. 

\subsection{Methods for Personalizing Assistive Technology}

In accessibility research, personalization of technology to meet user needs is used to reduce the burden of accessibility on users \cite{garrido2012personalized, sloan2010potential, stangl2021going}.
This typically leaves the function of the technology unchanged, but aims to automatically map the input and output mechanisms to new systems or modalities \cite{gajos2007automatically, wobbrock2011ability}.
For example, Yamagami et al. recently considered how people with motor impairments would create personalized gesture sets that map to common input mechanisms \cite {yamagami2023people}.

Work customizing the functionality of assistive technology is more limited in comparison.
In AI assistive technology, teachable object recognizers have been used to allow users to personalize recognition models themselves \cite{kacorri2017people}. 
Users capture their own image or video data of unique objects that can be stored and later recognized \cite{theodorou2021disability, findmythings}. 
These approaches can be more useful than off the shelf object recognition models as they are customized to user's specific needs \cite{kacorri2017teachable, bragg2016personalizable}.
However, for commercial applications, users have limited avenues for customization. While screen readers can be personalized through a variety of settings, shortcuts, and add-ons \cite{momotaz2021understanding}, AI powered assistive applications are typically part of closed software ecosystems. While they may have some settings within the application for things like language and output speed, this is typically the extent of the customization.
Through this work, we hope to demonstrate new methods for personalizing assistive technology functionality to meet unique user needs.

\subsection{DIY Assistive Technology}

DIY communities have adopted an approach to making centered around personalization, democratization, and collaboration \cite{kuznetsov2010rise, tanenbaum2013democratizing}. 
For assistive technology, DIY approaches can help to address assistive technology adoption due to unique or changing needs \cite{hurst2011empowering}.
To this end, prior research on DIY assistive technology has sought to make the process of prototyping and making more accessible to participants with a range of technical skills \cite{meissner2017yourself, rajapakse2014designing}.

Most of this research focuses on making physical tools for accessibility (e.g., making 3D-printed devices like an ironing guide, right angle spoon, or tactile graphics \cite{buehler2015sharing}, prototyping custom prosthetics \cite{hofmann2016helping}), rather than software tools.
Some tools are being developed to support blind people in DIY-ing more high-tech hardware sensing systems, such as A11yBits \cite{he2023multi} or the Blind Arduino Project \cite{blind-arduino}. 
While these raise the ceiling of high-tech DIY creation, little research has focused on DIY-ing new software systems for existing devices users already own. In their original case studies of DIY assistive technology, Hurst and Tobias highlighted one instance of `high-tech custom-built assistive technology', wherein a team of professional programmers worked with an artist with ALS to create software that used eye-tracking input for drawing \cite{hurst2011empowering}.
Lowering the barrier to entry for creating technically complex assistive software is an important next step in enabling people to DIY personally meaningful assistive technology.

\subsection{End-User Programming}

Decades of end-user programming research has sought to understand and support programming work done by people who are not trained as programmers \cite{myers2006invited}. While initially focusing on end-user programming in professional contexts (e.g., using spreadsheets or other domain-specific tools \cite{scaffidi2005estimating}), a variety of approaches have been developed to support programming for personal utility as well.
For example, Marmite is an end-user programming tool that allows users to create new applications by combining data and services from multiple existing websites \cite{wong2007making}.
Here, we describe previous end-user programming approaches that work towards the goal of making programming more approachable for novices.
In this work, we aim to apply these existing end-user programming approaches to the domain of visual assistive technology, enabling blind people to have a new level of control over assistive software.

\textit{Block-Based Programming.}
Visual, block-based programming approaches allow users to create programs by graphically organizing elements. These approaches often aim to support novices by providing pre-structured statements to reduce or eliminate syntax errors \cite{mohamad2011block}, for example, as in Scratch \cite{scratch}. While these approaches are commonly used in educational settings \cite{weintrop2019block}, they are also used in commercial mobile automation tools to provide sets of components that users can arrange as they wish to create time-saving automations, as in Shortcuts on iOS \cite{apple-shortcuts} and Google Assistant \cite{google-routines}.

\textit{Natural Language Programming.}
Further work has aimed to synthesize programs from natural language alone.
These approaches commonly require a set of training data consisting of queries and desired automations \cite{desai2016program, li2020mapping}.
Large language models have also been used for program synthesis, with mixed results \cite{austin2021program, vaithilingam2022expectation}.

\textit{Programming By Example.}
Programming by example approaches alternatively allow users to create programs by providing demonstrations of desired functionality, without the need for any code \cite{lieberman2001your, cypher1993watch}.
Programming by example has been implemented in a range of domains, for instance, Rousillon automates web scraping with a demonstration from users on how to collect the first row of a data table \cite{chasins2018rousillon}, and Sugilite automates actions on mobile interface using a demonstration and natural language request \cite{li2017sugilite}.

\textit{End-User Programming and Accessibility.}
The accessibility of programming tools is a nascent area \cite{potluri2022psst, potluri2018codetalk, pandey_accessibility_2022} that has largely focused on developers rather than end users.
End-user programming approaches have occasionally been applied to accessibility contexts for the purposes of sharing accessibility bugs and teaching blind children.
For example, for web accessibility, demonstration has been used as a method for end-users to convey accessibility issues to developers
\cite{bigham2007accessmonkey, bigham2010accessibility}.
Story blocks is a tangible block-based tool for teaching blind students programming concepts \cite{koushik2019storyblocks}.
We aim to continue and extend this line of research by supporting blind end-user programmers in creating new visual assistive software.

\begin{figure*}[t!]
 \includegraphics[width=1\textwidth]{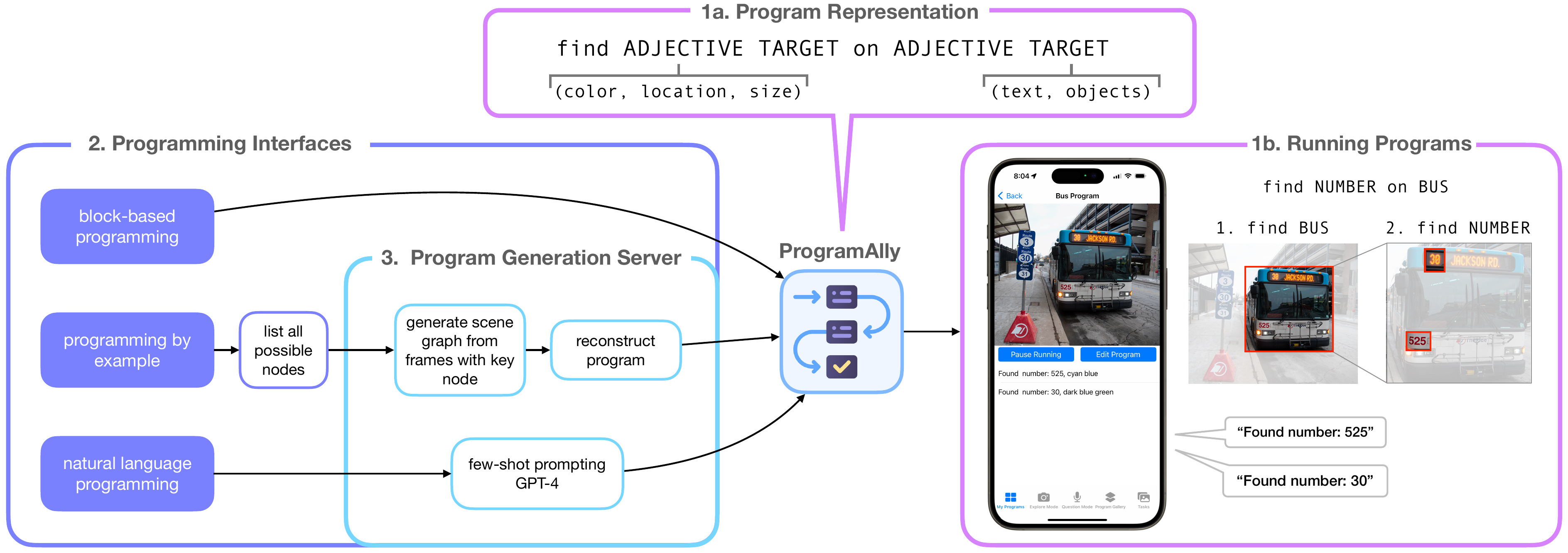}
 \vspace{-1pc}
 \caption{\sys's main components: (1a) An underlying program representation, the framework for running visual filtering programs (1b). (2) A set of three, multi-modal programming interfaces to support programmers with different levels of expertise. (3) A program generation server which synthesizes filtering programs from images or natural language.}
 \Description{Figure 2: ProgramAlly system diagram showing the 3 main components: (1) program representation for creating and running programs, (2) program creation interfaces, and (3) a program generation server. (1a) The underlying program representation is in the form 'find ADJECTIVE TARGET on ADJECTIVE TARGET' with adjectives being color, location, or size, and targets being types of objects or text. (1b) Shows a diagram of how programs are run. For the example program 'find NUMBER on BUS', the app first looks for buses in the frame. Then, if a bus is found, the image is cropped to just contain the bus. In the cropped frame, text detection will be used to look for numbers. (2) The app has 3 ways to create programs: block-based programming, programming by example, natural language programming. (3) The program generation server is used to generate programs from the latter two interfaces. For programming by example, the app lists all possible nodes/ items to filter for. Then the server generates a scene graph from frames that contain that selected node, and uses that graph to reconstruct the program. For the natural language programming mode, the app uses a few-shot prompting approach for GPT-4.}
 \vspace{-0.75pc}
 \label{fig:sys-diagram}
 \centering
\end{figure*}

\section{\sys}

\sys is a mobile application that implements end-user programming techniques to allow users to create block-based visual information filtering programs (e.g., \texttt{`find NUMBER on BUS'}). 
\sys is implemented as a native iOS application, and consists of the following components, as shown in Figure \ref{fig:sys-diagram}:
\begin{enumerate}
 \item \textbf{A program representation}, as the framework for implementing and running programs with on-device models.
 \item \textbf{Program creation interfaces} provide a multi-modal set of tools for users to create and iterate on programs.
 \item \textbf{A program generation server} provides the components for automatically generating programs based on images or natural language text.
\end{enumerate}

\subsection{Design Goals}

Overall, we designed \sys based on three primary goals: Expressiveness, Approachability, and Accessibility.

\textbf{D1: Expressiveness.} \sys's goal is to be an interface where users can customize off-the-shelf models for their own uses. Programs should be able to support a wide variety of real-world use cases through a flexible structure and range of models.

\textbf{D2: Approachability.} \sys needs to be approachable for non-experts. To this end, it includes a set of methods to create and iterate on programs through multiple modalities, and users can choose what fits their needs. 
\sys should aim to have as little technical jargon as possible and explain program parameters in natural terms.

\textbf{D3: Accessibility.} \sys needs to be VoiceOver and Braille display accessible for users, both while creating and running programs. \sys's VoiceOver implementation groups related parameters together to provide context for each statement. Additionally, \sys provides visual context while running programs to help the user aim and know what is in frame.

\subsection{Visual Filtering Programs in \sys}

\sys is built on a generalizable representation of visual filtering tasks. Here, we describe how that represntation was designed, how it is implemented, and how it is used to run visual filtering programs.

\subsubsection{Designing a Representation of Filtering Tasks}
\label{sec:filtering-design}

\sys's scope of programming visual filtering tasks was determined based on prior work indicating it to be a possible domain for assistive technology customization \cite{herskovitz2023hacking}. 
We aimed to understand features of filtering tasks in order to build a program representation that could capture a variety of user needs (D1: Expressiveness).
Herskovitz et al. captured a dataset of scenarios where blind participants described cases of wanting to create or customize assistive technology \cite{herskovitz2023hacking}.
From this dataset, we labeled specific instances as filtering tasks: cases where the participant was searching for a certain type of information.
We considered a task to require filtering if using a general scene description tool like Be My AI \cite{bemyai} or OCR like Seeing AI's Document Mode \cite{microsoft-seeingAI} would produce extraneous or distracting information beyond the intended task, but could produce useful results with additional processing.

Out of the original set of 201 scenarios, we identified 29 as filtering. These were fairly evenly spread across all 12 participants from the dataset, with each participant describing at least one.
Scenarios fell roughly into two types: finding specific types of text, or specific items.
For searching for text, this could be finding specific strings (i.e., a name on a package, a room number in a hotel), finding certain types of text (i.e., a number of miles on a treadmill, the number of calories on a package), or finding text in a specific location (i.e., on a thermostat display, on a license plate).
For searching for objects, this could be finding a specific type of item (i.e., a trash can in a mall, a stairway), or items in a specific location (i.e., a person in a chair, an obstacle on a sidewalk). 

From this analysis, we determined two key aspects to include in our filtering program representation: (1) the ability to filter by type of object or text, and (2) the ability to filter by an item's location. Our representation includes two types of statements to address this: a `find' statement, and an `on' statement to convey objects overlapping. 
We confirmed this representation by analyzing a random sample of questions from the VizWiz Question Answering dataset, a dataset of images and questions asked by blind people \cite{gurari2018vizwiz}.
We found that the two statements in our representation could represent a significant portion (approximately half) of the 100 queries we analyzed, without the need for additional operators that would increase program complexity.

\subsubsection{Program Representation}

Programs in \sys are generally in the form \texttt{`find ITEM on ITEM'}, with any number of `find' or `on' statements. For example, a program can range from \texttt{`find CAR'} to \texttt{`find COLOR on CAR'} to \texttt{`find TEXT on LICENSE PLATE on CAR'}. Adding multiple `find' statements runs each `find' statement in parallel and produces a similar effect to an \texttt{OR} operator.
For example, the program \texttt{`find COLOR on CAR, find TEXT on LICENSE PLATE on CAR'} for locating a ride share would announce both the color and license plate number of a car if visible.

Additionally, each item in the statements can consist of both a target item (e.g., an object, a type of text), and an optional adjective to describe that target. \sys supports adjectives denoting color, size, or location. For example, \texttt{`find NUMBER on RED BUS'}, \texttt{`find LARGEST TEXT on SIGN'}, or \texttt{`find ADDRESS on CENTER ENVELOPE'} are programs where the output would be further restricted to match specific conditions. Programs in \sys are stored as lists of these items (adjective and target pairs). 

\subsubsection{Running Programs}

\sys uses this representation to run programs to generate live output.
\sys does this by iterating over the list of items in a program backwards, cropping or filtering the source image at each step. 
For example, in the \texttt{`find NUMBER on BUS'} program shown in Figure \ref{fig:sys-diagram}, \sys first runs an object detection model that has the class `bus'. The model will output a series of bounding boxes that have that class label. 
Then, the next item in the program is processed. In this case, for each bus bounding box, the frame is cropped and passed into a text detection model. The resulting text is then filtered for strings that only consist of numbers.
\sys keeps track of points where filtering fails (say, if no buses are found), and later uses that information to generate descriptive program output.

\sys currently leverages a set of models and functions for processing each piece of a program. 
Each item that can be included in a program is stored in a dictionary associating it with the relevant model.
The two primary types of targets, objects and text, are both recognized with a set of on-device models, though this could be extended in the future to support cloud models as well.
For object detection, \sys uses a set of YOLO models as they have low latency on a range of iPhones.
This includes the default set of 80 object classes detected by YOLOv8 \cite{reis2023realtime, yolo8}.
Additionally, we included a set of modified YOLO-World models, a version of YOLOv8 that can be extended with new detection classes without any fine-tuning through a vision-language modeling approach \cite{cheng2024yolow, yolo-world}.
We added four additional models with classes that we chose to be relevant to accessibility tasks (D1: Expressiveness) \cite{brady2013visual}: an outdoor navigation model (`sign', `license plate'), an indoor navigation model (`door', `stairs', `hallway', `exit sign', `trash can'), a reading model (`envelope', `package', `document'', `poster'), and a product identification model (`package', `can', `bottle', `box', `product', `jar').
For this last set of classes, we made them available under one super-class called `grocery item' for flexibility.
New models can easily be added to \sys as it searches the dictionary for the appropriate model when running a program.

\add{In addition to recognizing objects, \sys also detects text with iOS’s native text recognition. Programs can include the item `any text', but can also include various more specific types of text such as `address', `email', `phone number', `date', etc. These types can all be used within programs, for example, \texttt{`find ADDRESS on PACKAGE'}. These text types are detected by a combination of Google’s Entity Extraction API \cite{entity-extraction} and regex functions.
}

\add{Finally, adjectives in \sys are then used to further filter object or text results.
Adjectives include color (red, blue, etc.), size (largest, smallest), and location (center, upper left, etc.).
These can be used alongside any item, for instance in the program \texttt{`find LARGEST TEXT on BLUE SIGN'}. 
Adjectives in \sys were implemented natively:
color is detected by matching the most common pixel colors within an object's bounding box to a set of strings;
size is determined by comparing an object's bounding box to others of its type and then filtering for the lower or upper quartile;
and item location is determined based on a quadrant system, breaking down the parent object (either the image frame or a bounding box) into sections to label the location of a child item (i.e., ``text on upper left'').
While these implementations are naive, they are meant to demonstrate that a variety of sources of classification can be used in \sys, and could eventually be replaced by more robust models or algorithms.
}

\subsubsection{Program Output}

While running programs, \sys keeps track of where target items were found in order to give context for each piece of information. 
For example, if two buses are found in the frame, the output could be: ``Found number 73 on bus, left of frame, found number 21 on bus, right of frame.'' 
This system also tracks where the program failed if the target was not found. 
If the first item in the program is not found, \sys will attempt to provide output for the second item, and so on. For example, if a bus is found with no number on it, the output would be, ``Found bus, no number.''
In this case, because `number' results are filtered from the more general text detection model, \sys would also read strings that are not numbers as a backup, for example, the route name.
Unique messages are generated for each failure point, for example, if no buses are found (``No bus found''), or if a bus is found but the adjective does not match (``Found white bus, no red bus visible'').
This information is used to provided helpful backup information for understanding the scene and aiming the camera.

\subsection{Block-Based Programming Mode}

\sys's first method for creating new filtering programs is a block-based programming interface, shown in Figure \ref{fig:teaser}.
This block mode can be used to create a program from scratch, or to edit a program that was generated automatically by one of the other two methods.
When first creating a new program, the authoring interface will display a program with two empty items in order to provide a default structure for users to fill in.
There are two sections of this interface.
First, a heading called `Program Summary', which includes a natural language summary of the implemented program for users to refer back to as they edit. For the default program, this will initially read, ``Find any object on any object'', and will update as users fill in the program with their desired items.

Next, under a heading called `Edit Program Directly', users can read through each statement in the program, and edit them with actions in VoiceOver.
For example, when VoiceOver focus is on the first `find' statement, it will announce: ``Find any object, actions available: Edit adjective, Edit object, Delete this item.'' If parameters for the item have already been selected, the VoiceOver description changes to reflect what has been chosen. For example, for the statement \texttt{`find RED BUS'}, the description would be: ``Find red bus, actions available: Edit adjective `red', Edit object `bus', Delete this item.''
Grouping these together as one single element with multiple actions, rather than having `edit adjective' and `edit object' as separate VoiceOver elements, is meant to clarify that the different parameters in the `find' statement are functionally related, without relying on the visual aspect of them each being on one line (D3: Accessibility).
This design was also based on the VoiceOver experience of Apple's Shortcuts app \cite{apple-shortcuts}, where each block is read as a separate element and editing parameters can be similarly accessed through actions.
When either of the edit actions are activated, a new page will appear listing the possible adjectives or objects to fill in the program (shown in Figure \ref{fig:teaser}). 
The menu includes buttons that can be used to filter the items by type, or a search bar for finding a specific item.

\subsection{Natural Language: Question Mode}

Inspired by natural language programming approaches, \sys includes `Question Mode', which generates a program from a question or statement (D2: Approachability). Users can type or dictate a question, and the resulting program will appear in the block-based interface for them to review and refine further. 
\add{For instance, the query, `What does this bottle say?' would result in the generated program: \texttt{`find ANY TEXT on BOTTLE'}. This result could then be modified with a follow up question: `Actually, just read the biggest text' changes the program to \texttt{`find LARGEST TEXT on BOTTLE'}.
}

To prototype this interaction, we use a few-shot prompting approach with GPT-4. We provide a custom system prompt describing how to extract items, and listing the possible item classes. Then, we provide a set of approximately 20 queries and their correct JSON program representation that we wrote based on examples from accessibility datasets \cite{herskovitz2023hacking, gurari2018vizwiz}. 
Without developing a custom entity extraction workflow, we found that this approach works well. However, GPT-4 will sometimes produce errors. The most common issue is the model hallucinating new object classes. In this case, the block interface will alert the user that there is an unsupported field and open the editing menu for users to select an alternative. The model very rarely produces programs with an incorrect structure. If the model fails to extract entities, which can happen if the question is vague (e.g., ``What is this?'') it will occasionally respond with natural language rather than a program (e.g., ``I'm sorry, I don't know what you mean, can you clarify?'').
Future work could use additional fine-tuning to create a conversational approach to clarifying ambiguous language.

\sys also includes a method for users to edit programs with a follow-up question, rather than by manually editing a generated program using the block-mode. When editing a generated or pre-existing program, there is a text box where users can type or dictate a follow-up question (Figure \ref{fig:teaser}). Various follow up questions are included in our system prompt to GPT.
Based on feedback in our formative studies, we also included an option to access this feature while a program is running. If the program output is not as expected, users can directly access the option to edit the program with natural language, for rapid iteration. For example, when running the program \texttt{`find NUMBER on BUS'}, the user could provide the statement, ``Read the route name instead'', and the program would be modified to be \texttt{`find TEXT on BUS'}. 

\subsection{Programming-By-Example: Explore Mode}

\begin{figure}
 \centering
 \includegraphics[width=0.95\linewidth]{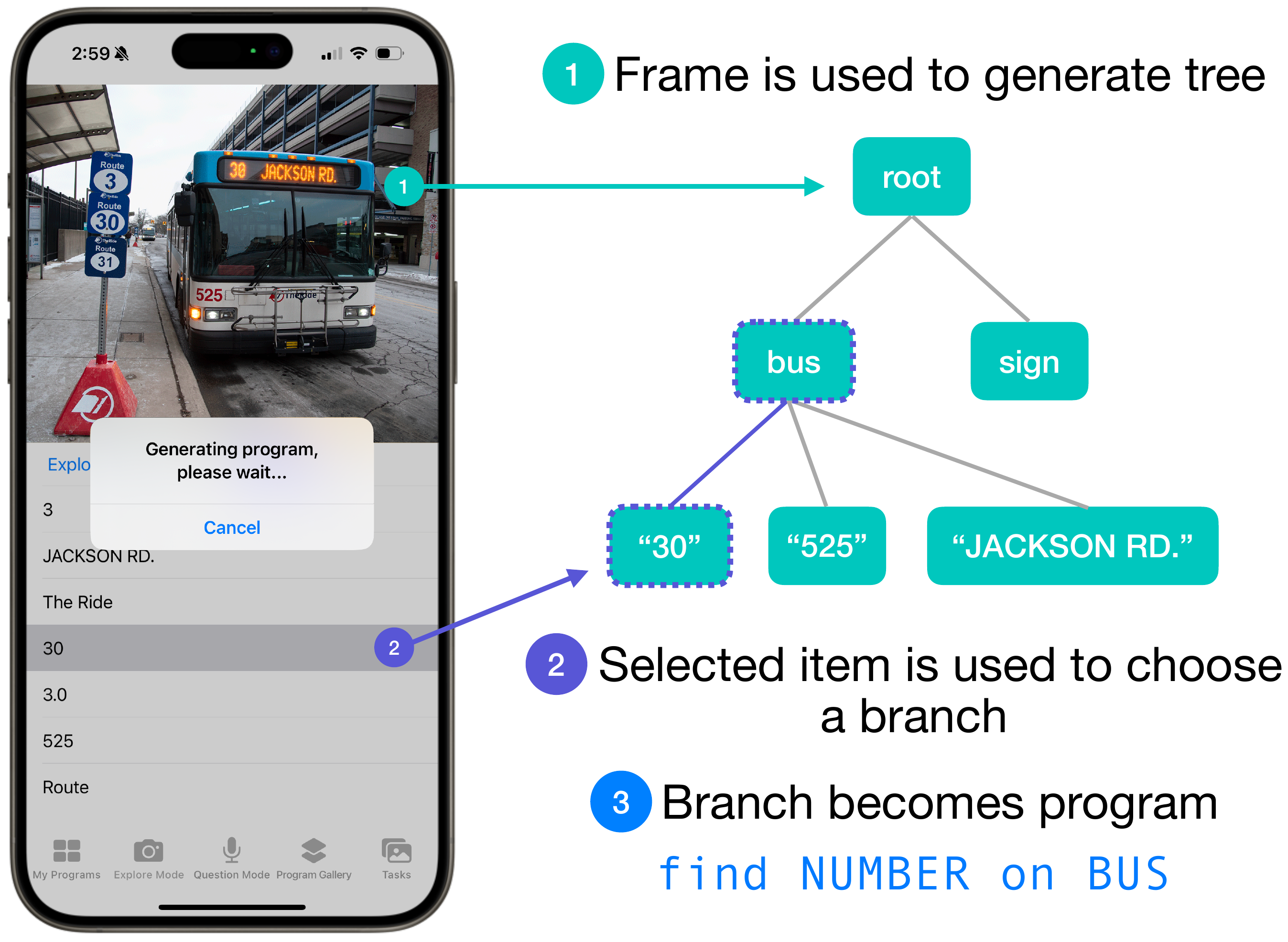}
 \caption{In explore mode, \sys provides a list of all items detected in the camera feed. Users then demonstrate filtering by choosing a specific item. That item is then used to fetch a specific branch from a scene hierarchy, which becomes the program.}
 \Description{Figure 3: iPhone screenshot of showing the interface of explore mode, as it is processing an image of a bus. The selected item is text that reads '30.'
There is a diagram showing how the program is generated. First, the frame is use to generate a tree containing visual features in the image like “bus,” “sign,” “30,” “525.” They are arranged in a hierarchy, so the first level of the tree is the root frame, with the children being 'bus' and 'sign'. Then, the bus has child elements for the text on it.
When the '30' is selected, because this is a child element of 'bus', that branch from the tree is used to generate the program “find NUMBER on BUS”.}
 \vspace{-1pc}
 \label{fig:explore-mode}
\end{figure}

\sys's Explore Mode allows users to automatically generate a program by selecting a target feature detected in the camera feed (D2: Approachability).
Explore mode lists all object and text features in the image, and users select an item to filter for, effectively providing a demonstration of the filtering behavior.
Explore mode was included to address the challenge of unknown-unknowns \cite{bigham2017effects}: without knowing what visual features or information is present, blind users would not necessarily have all of the information available to write a working filtering program.

In this mode, \sys runs all object and text detection models at once, with the goal of outputting everything that a user might want to create a program to find. Users then select the information they are looking for to demonstrate filtering behavior, and a program is generated which aims to filter for that type of information in the future. For example, as shown in Figure \ref{fig:explore-mode}, the camera is pointing at a bus stop. If the user selects `30', which is a route number, the resulting generated program will be \texttt{`find NUMBER on BUS'}, because that is where the text `30' was found in the frame.
Once a program is generated, the app again displays the result in the block-based interface, with the program summary and the option to edit the generated program further either with a question or with blocks.

\subsubsection{Generating Programs from Demonstrations}

Programs are generated using \sys's server, which represents images as a tree of items.
First, on the device, \sys maintains a list of each frame where an item was detected. When the user selects an item, a frame is then chosen that contains that item, which is then sent to the generation server.
\sys's server uses a set of models to then generate the tree structure. These are slightly different than the models used on device, and includes Mask R-CNN under Detectron2 \cite{detectron} and Google's Cloud Vision API \cite{google-vision} for object detection and Google Cloud's OCR model \cite{google-ocr} for text.
Additionally, to label adjectives and other properties associated with each item, the server runs DenseCap \cite{densecap}, a model that creates rich language descriptions of image regions.
Because DenseCap produces natural language descriptions associated with bounding boxes, we use a few-shot prompting approach to GPT-4 \cite{gpt4-report} to extract and label the relevant objects and their associated adjectives.

Next, each item is stored as a node in a scene graph hierarchy based on bounding box overlap. The parent node is the entire image, and child nodes can either be text or objects, stored with their associated adjectives. 
Finally, from this scene graph, the originally selected node is then used to generate a program. The selected node is located in the scene graph, and all of its ancestor nodes (not including the root image) are then selected, representing a branch of the graph (see Figure \ref{fig:explore-mode}). Traversing up this branch, each node then becomes an item in the program. Each node in this set is converted into an adjective and object pair, and ordered based on their parent-child relationship in the source graph. This generated program is then sent to the device as JSON.
While \sys currently uses a strict tree structure to avoid any ambiguity (ensuring that a single branch can always be chosen), this does limit this generation technique to supporting only the current `find' and `on' operators. To support more complex programs, new synthesis techniques would need to be developed.

Because the server includes the addition of DenseCap for describing objects, there may in rare cases be a class in the generated program that is not present in the app, although we aim to filter these classes out when possible. 
In this case, the block interface will again alert the user that the field is unsupported and surface the menu for selecting a replacement. 
\section{User Study Protocol}

\begin{table*}
 \renewcommand{\arraystretch}{1.1} 
\centering
\small
\begin{tabular}{l|l|l|l|l}
\textbf{ID} & \textbf{Gender} & \textbf{Age} & \textbf{Occupation} & \textbf{Vision and Hearing Level} \\ \hline
P1 & Man & 22 & Computer science university student & Blind with no light perception, from age 5 \\ \hline
P2 & Woman & 54 & Not employed & Blind with some light perception, from age 49 \\ \hline
P3 & Man & 29 & Graduate student & Blind with some light perception, from birth \\ \hline
P4 & Woman & 32 & Program Director & Blind with some light pereption, from birth \\ \hline

R1 & Woman & 41 & Translator & Blind, from birth \\ \hline
R2 & Man & 37 & Accessibility consultant & Blind, from birth progressive up to age 15 \\ \hline
R3 & Woman & 68 & Retired & Blind with some light perception, progressive since birth \\ \hline
R4 & Man & 41 & Assistive technology research and training specialist & Blind with no light perception from birth. Deaf/ hard of hearing. \\ \hline

F1 & Woman & 53 & Not employed & Blind with no light perception, from age 51 \\ \hline
F2 & Woman & 60 & Retired teacher & Blind/ low vision, 20/500, from age 27 \\ \hline
F3 & Man & 40 & N/A & Blind with some light perception, from birth. Slightly hard of hearing. \\ \hline
F4 & Woman & 72 & Retired social worker & Blind with some light perception, from age 23 \\

\end{tabular}
\vspace{0.5em}
\caption{Participant demographics for our study with 12 visually impaired people. Participants self-described their level of vision. All participants used a screen reader to access their devices and read text. Participants with a `P' were part of the pilot testing, participants with an `R' completed the full study remotely, and participants with an `F' completed the full study face-to-face.}
\vspace{-1pc}
\label{tab:participants}
\end{table*}

To understand how \sys can be used as a tool for creating and customizing assistive technology, we conducted a study with 12 blind participants. \textbf{Our goals were to (1) assess the accessibility and approachability of \sys, and (2) understand unique challenges faced by blind end-user developers creating visual technology.} 
This study was approved by our Institutional Review Board (IRB). Participants were compensated \$25 per hour for their time and expertise. This ranged from 1.5 to 3 hours in total, with an average time of 2 hours.

We aimed to involve participants in \sys's design, so the first four participants were consulted as it was being developed and informed many of its final features. 
Because these participants also completed a similar study protocol as the remaining participants, we include their results here as well.
Overall, this study was completed with three groups of participants:
\begin{enumerate}
 \item \textbf{Pilot Participants:} Four remote participants who tested \sys as it was being developed. The first two participants only used the block-based programming mode, and the second two participants used all three modes.
 \item \textbf{Remote Participants:} Four remote participants who completed a full evaluation of \sys, running filtering programs on sample images.
 \item \textbf{Face-To-Face Participants:} Four in-person participants who completed a full evaluation of \sys, running filtering programs on provided props and comparing filters to existing assistive apps.
\end{enumerate}

\subsection{Participants}

Participants were recruited using email lists for local accessibility organizations, prior contacts, and snowball sampling. Participants were required to be over 18 years old, have some level of visual impairment, and regularly use a screen reader to access their devices. Participants were also required to have an iPhone so that they could download \sys via TestFlight.

Demographic information for participants is shown in Table \ref{tab:participants}. Of the 12 participants, two had some prior programming experience for their coursework or career. However, participants had a range of experiences with technology and VoiceOver, and not all were experts. For example, R2 and R4 were assistive technology professionals, while F2 was new to using VoiceOver and had not previously used any mobile assistive applications.
We recognize that recruiting remote participants can create a bias for people who are technically savvy, as they need to have a desktop and be familiar with Zoom. 
To try diversifying our sample, we also recruited in-person participants.

\subsection{Procedure}

\begin{figure}
 \centering
 \includegraphics[width=1\linewidth]{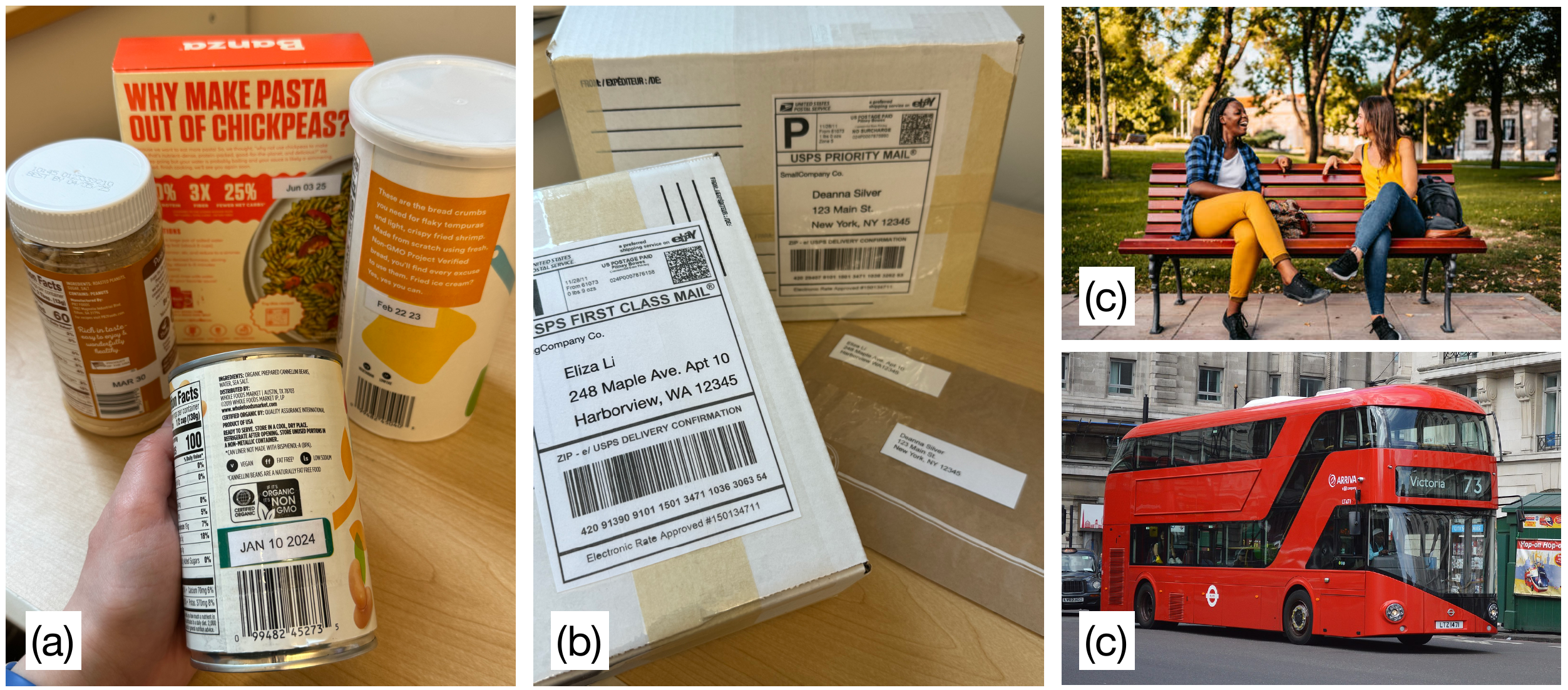}
 \vspace{-1.5pc}
 \caption{Samples of props used in our study: (a) Grocery props for in-person testing of \texttt{`find DATE on GROCERY ITEM'}, (b) Mail props for in-person testing of \texttt{`find ADDRESS on PACKAGE'}, (c) Images used by remote participants, for testing \texttt{`find NUMBER on BUS'} and \texttt{`find PERSON on BENCH'}.}
 \vspace{-1.3pc}
 \Description{Figure 4: Sample props. (a) cans and a pasta box. (b) mail package with a printed UPS label, two more packages in the background. (c) two sample images, on top an image of two people sitting on a park bench, on the bottom a red double-decker bus.}
 \label{fig:study-tasks}
\end{figure}

After a brief introductory interview, participants were introduced to \sys by reading through a pre-written program to familiarize themselves with the concept and interface. Participants were then asked to modify the example program slightly by adding an adjective.
We tried to keep verbal instructions minimal to let participants reason for themselves about the program.
The only pointer that we gave to participants was that they could swipe up or down on the program elements to hear the different editing actions available, because depending on the verbosity settings of their device, VoiceOver may not have spoken this information.

After this introduction, participants then used \sys to create and run three programs.
Participants used each of the three programming interfaces (block mode, explore mode, question mode), and were assigned tasks to create a program for (i.e., `create a program that will find addresses on mail').
While creating each program, participants were prompted to think aloud to explain their thoughts or concerns, or to ask questions.
Participants then ran the programs they created, with remote participants using sample images and in-person participants using props. 
These props included books, grocery items, and packages and mail, and examples are shown in Figure \ref{fig:study-tasks}.
In-person participants were also given the option to compare their program to either Be My AI or Seeing AI, depending on what they would normally use for the same task.
After creating each program, participants were asked to rate the ease of creating the program, and how accurate they felt the program was.

Finally, participants were asked to rank the three creation modes on different factors. The study concluded with an open-ended interview about the prototype app and the experience of using end-user programming methods for DIY assistive technology overall. 
Participants were asked to imagine how such an app could fit into their existing assistive technology workflows, and the pros and cons of creating programs to customize assistive technology.

\subsection{Data Collection and Analysis}

Remote participants joined a Zoom call that was recorded, and when testing the app they were asked to share their phone's screen in the call.
Similarly, the face-to-face participants interviews were audio recorded, and their devices were screen recorded.
The audio was later transcribed and used for analysis.
We then created written descriptions of participant's strategies for completing each tasks from the video data.

Since participants were encouraged to use a think-aloud method and take their time to fully explore the functions, ask usability questions, and give feedback, performing an analysis of task completion time does not provide much insight about how \sys works in practice.
Instead, we primarily report qualitative data on participants’ strategies and workflows while using \sys, and their general feedback.

\section{User Study Results}

Here we present results from our user study. 
First, we discuss participants' experiences creating and running programs in general.
Next, we discuss how participants used and compared each of the three program creation modes.
Finally, we discuss participants impressions of end-user programming as a customization tool.

\subsection{Using Filters in \sys}

Participants generally felt positively about using filtering programs in \sys. As P1 described: \textit{``There is an abundance of visual information. And sometimes a blind person is short on time, and you really just want the particular piece of information you're curious about''} (P1).
Not all of the example filtering programs were equally useful to participants.
For example, R1 appreciated the \texttt{`find PERSON on BENCH'} program and envisioned using it in a local park with many benches, while R3 had a seeing eye dog already trained to do this task.
All participants saw practical use in at least one of the filters they tried, and all were able to come up with ideas for filters they could create in the future.

\subsubsection{Benefits of Filtering}

Almost all participants saw filtering as an important niche not filled by existing assistive technology.
As R3 described: \textit{``I definitely see a use for this app. Earlier I mentioned that I usually think, `I already have 3 apps that do the same thing. Why do I need another one?' But this one, because of the ability to filter, you know, as much as you want to, you can get very specific, I don’t think anything like that exists. Or maybe it exists in a 5 step process''} (R3).
Similarly, R4 mentioned how filtering could speed up some tasks: \textit{``I think it’s the next natural direction to go in, in some ways, sometimes you get a bit too much information and you can speed up the process by not having it automatically generate an image description that you might not even have any use for''} (R4).

Participants expressed that filtering seemed particularly useful for tasks that were repeated as part of their routines.
F2 said, \textit{``I probably would save some [programs] to the library. Because some stuff I would probably use over and over again''} (F2).
In routine tasks, efficiency can be more crucial.
F3 said, \textit{``Well, it depends on your routine, you know. If you have something that you do on an ongoing basis, and you really need this life hack, say, to make it real simple for you, then absolutely''} (F3).
Participants were also able to envision creating new filters outside of the ones prompted in the study. A list of examples is shown in Table \ref{tab:examples}.

\begin{table}
 \renewcommand{\arraystretch}{1.1} 
\centering
\small
    \begin{tabular}{p{0.08\linewidth} | p{0.84\linewidth}}
        \textbf{ID} & \textbf{Program Idea} \\ \hline
         P1, F2 & Identify products by brand or flavor when sorting, \texttt{find BRAND NAME* on GROCERY ITEM}\\ \hline
         P3 & Find cooking instructions, \texttt{find TEXT AFTER "COOKING INSTRUCTIONS"* on GROCERY ITEM} \\ \hline
         P3 & Sort credit cards, \texttt{find NAME* on CREDIT CARD*} \\ \hline
         R3, F3 & Organize books and CDs, \texttt{find "AUTHOR NAME"* on BOOK}\\ \hline
         R4 & Find room number in hotel, \texttt{find NUMBER on SIGN} \\ \hline
         F3 & Identify car models, \texttt{find MAKE/MODEL* of CAR} \\ \hline
         F4 & Differentiate between two cats, \texttt{find ORANGE CAT} \\
    \end{tabular}
    \vspace{0.5pc}
    \caption{Participants envisioned creating new filtering programs outside of those they used in the study, sometimes involving new classification models (marked with a *). A sample of these ideas are shown here.}
    \vspace{-2pc}
    \label{tab:examples}
\end{table}

\subsubsection{Comparing \sys to Other Assistive Technology}

In-person participants were able to directly compare filtering programs they made to other automated assistive apps of their choice.
In general, participants preferred using the filters they wrote over Seeing AI for the same task. For example, F2 reflected on the \texttt{`find ADDRESS on PACKAGE'} program: \textit{``The filter, I think, did center just on the address. It read it one time, and I knew where the start and finish was. In Seeing AI, it seemed like it would keep reading and reading and reading... I would have to listen to it twice to make sure I got the full address, So that's why I slowed it down, because I really had to pay attention where the start of the address was and what it was telling me''} (F2).
Here, when F2 used Seeing AI to read a shipping label, it read additional extraneous information aside from the address that they then had to manually sort through, to the point where they reduced VoiceOver's speaking rate to listen carefully for the information.
Similarly, F1 and F3 preferred the \texttt{`find DATE on GROCERY ITEM'} program over Seeing AI, as Seeing AI never read an expiration date, it just read other information on the product package, despite the date being in frame.

Participants preferred Be My AI for some tasks, but not all of them.
For example, F4 compared the \texttt{`find DATE on GROCERY ITEM'} to Be My AI. After trying multiple times, they were unable to take a photo for Be My AI with the expiration date in frame, and gave up.
On the other hand, participants tried a filter \texttt{`find LARGEST TEXT on POSTER'}, which was created imagining a scenario where someone would want to skim over fliers on a bulletin board.
Participants though Be My AI was better suited to this task, as it was easy to feel a flyer to center it in the frame, read the first line of the output description, and disregard the rest. 
This confirms our hypothesis that \sys is better suited to tasks that are continuous or repetitive, where taking a single photo is difficult.

\subsection{Programming Process and Challenges}

Most participants felt like with practice, they could become more familiar with the system and would be able to quickly create new programs.
However, as blind end-user developers they also faced unique programming challenges.

\subsubsection{Programming with Unknown Unknowns}

One such challenge is not knowing what the `ideal' program would be due to not knowing the parameters or targets included in the system.
While this is a challenge for many end-user programming tools \cite{ko2004six} and can be alleviated with more familiarity, R3 pointed out that this is also tied to prior visual ability:
\textit{``I'm gonna point out that, depending on your onset of blindness, you might not know what questions to ask. So it's going to depend on the user and their life experiences... Because their experiences, haven't you know, given them the ideas of how to word a question''} (R3).

To potentially address this, R3 imagined follow up information in the example and question modes that would help them understand the possible programs better. They said, \textit{``If there were things about this object that I wanted to know, but I didn't know that they're there, then there are questions that I'm not coming up with that would help me. So you know, in that respect, if the app gave me suggestions like, `Why don't you ask it this?' It might help you get to your final question''} (R3).
Explore mode in particular could be improved by listing possible programs, rather than only listing possible features of interest.

\subsubsection{Understanding Object Classes}

Even when participants were aware of a possible object class, they often wondered about the extent of what it would detect.
For example, P1 questioned the program \texttt{`find DATE on BOTTLE'}: \textit{``I don't know if it's super restricted to a specific definition of a bottle, or pretty much any box or container. Could it also work with a box of cereal, which also has an expiration date on it? A cereal box can hardly be, in colloquial terms described as a bottle. But from the AI's perspective, I wouldn't be too surprised''} (P1).
The ability to test classes in isolation, outside of a filtering program, could be helpful in determining the applicability or reliability of a class.

\subsubsection{Balancing Specificity and Reusability}

Participants thought carefully about how to produce filters that were specific enough to be useful, but generalizable enough to be re-usable.
For example, when writing the \texttt{`find DATE on GROCERY ITEM'} program, R2 noted that there may be both an expiration date and sell-by date on an item.
They said, \textit{``I guess with any of these filters is like, do you go broad, or do you go narrow? And I think `any date' works, with the knowledge that there are likely to be multiple dates available. That's not to say that I won't find the right information, but it just might find additional information that I didn't want''} (R2).
This is also somewhat dependent on prior visual ability, as someone with prior sight may be able to recall specific visual features that could cause conflicts in their filters.

Using only a single object detection class in a program could also limit its re-usability. For example, when reading the program \texttt{`find ADDRESS on PACKAGE'}, F4 noted they would want it to run on not just packages, but also on envelopes, mailers, or similar items. Creating new super-classes by grouping relevant items could improve the robustness of possible filters.

R4 noted that balancing program specificity could become more difficult when they were outside of their normal routine: \textit{``For example, in a situation where maybe I am trying to find a specific bus, I want to be able to quickly do that. So I would pre-write the program. But like, I could show up in a city next week, and not be there again for 3 years, so there would be no point in me generating different programs based on that. So it’s faster to just ask the question. But you know, in other situations, when I might have a little more time to put it together, or things like that, it seems like the other methods are a bit more accurate''} (R4).
Here, they described how if they needed a filter outside of one that they would typically use, they could quickly create one by asking a question, rather than spending time to carefully set up something they may only use a handful of times.

\begin{figure*}
 \centering
 \includegraphics[width=1\linewidth]{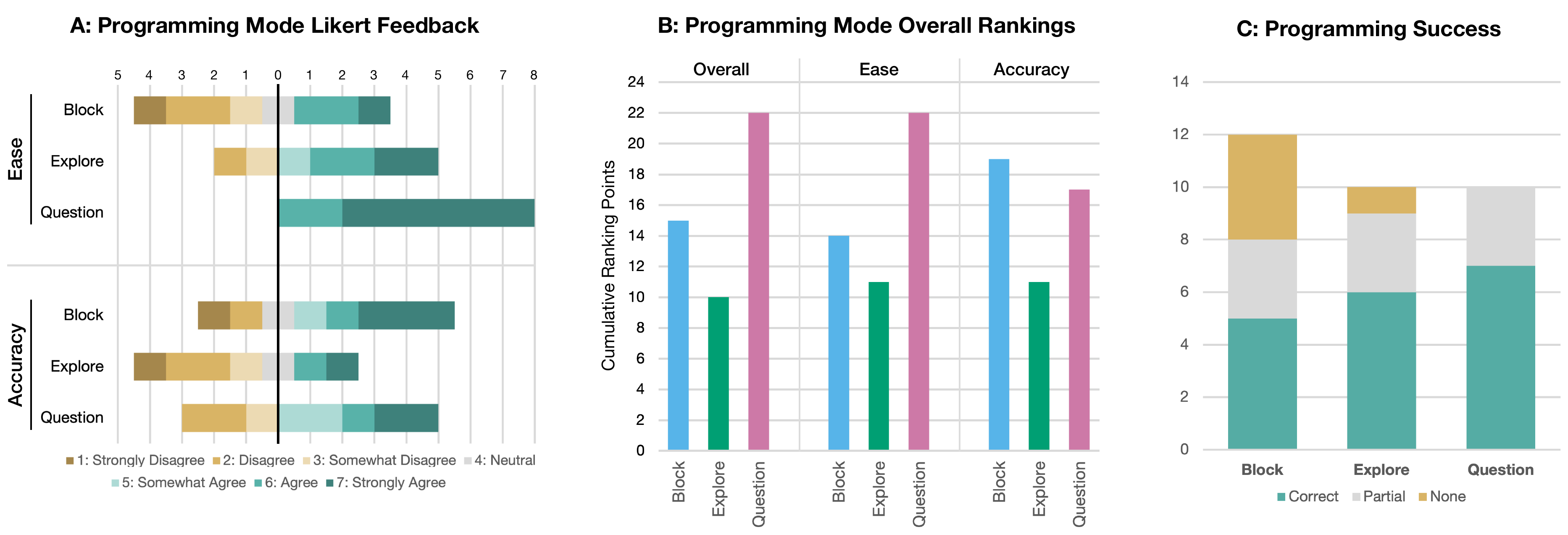}
 \caption{Participants rated each of \sys's three creation modes on a set of factors. Charts (A) and (B) demonstrate the trade offs between block and question mode: question mode was found to be easiest, but block mode was perceived to be slightly more accurate. Chart (C) demonstrates that block mode had the highest learning curve, though participants were able to create correct programs with all three modes. Each mode may be suited to different users or scenarios.}
 \Description{Figure 5: Three charts with different ways people rated ProgramAlly. (A) Programming Mode Likert Feedback. For ease, participants rated question mode as requiring the lease effort, all were affirmative that it was easy. Participants were splot on the other two modes. However, they rated block mode as slightly more accurate. Explore mode was seen as the least accurate. (B) Programming Mode Overall Rankings. Participants ranked each mode. Question mode was ranked first for overall favorite and easiest, but block mode was first for most accurate. Explore mode was lowest for all. (C) Programming Success. Shows how often people made correct or partially correct programs with each mode: 5 with block mode, 6 with explore, 7 with natural language. 3 participants were partially correct with each mode. 4 participants did not make a program with block mode, and 1 did not make a program with explore mode.}
 \label{fig:ratings}
\end{figure*}

\subsection{Creating Programs with Blocks}

\sys's block-based programming interface has the advantage of offering more fine-grained control over a program, but it also has the highest learning cost. 
However, participants expressed that this was something they felt confident in being able to learn over time as they created new filters.

\subsubsection{Advantages: Control and Accuracy}

Participants found that this mode produced programs that most accurately matched their intents because they could quickly specify exactly what information they wanted:
\textit{``I think this produces the most consistent accuracy, knowing if you can go in there and just select things''} (F1).
They also observed that this mode reduced some of the ambiguity present when a program was automatically generated for them:
\textit{``I would say it's the least difficult way, just because you know kind of what you're getting, and you know what's available''} (R2). Here, R2 is describing that when they already had a program in mind and knew the format, it was easier to program it directly than it was to think about how to phrase it in natural language.

\subsubsection{Learning to Think Programatically}

The block-programming mode required participants to adopt a programming mindset and break down an idea into multiple components.
For example, participants sometimes added items in the wrong order. 
P3 wrote the program \texttt{`find BUS on NUMBER'}, and described their thought process as \textit{``I was thinking, okay, first I need to look for a bus. That's the most important thing. So putting that first... I was going through a more like, linear process, you know, look for the bus then look for the number. But that's just like how the brain works. But if I had paid more attention to what it was saying...''} (P3). 
R2 also noted that it took some extra thought to consider how the scene would visually appear:
\textit{``For me, I was just thinking about it in terms of like, if you're looking for two objects which is going to be visually larger and easier to discover''} (R2). 
They then realized that they had put the elements in the wrong order when they went back to read the program summary: 
\textit{``This particular field helps me to determine which order I should put things. So I see that I said, `find any bench on any object'. So I actually want to find `any person' first''} (R2).

Additionally, when parameterizing their request participants were sometimes unsure if an item they wanted to add would be considered an adjective or an object.
R3 wanted to add the adjective `red' to an existing statement that said `find any book', but they selected `edit object' because they wanted to edit how the book was detected: 
\textit{``Cause I was like, `adjective'? I mean, I know what an adjective is, but I wasn't like relating it to `red book'''} (R3).
Similarly, F3 described, \textit{``It's like I have to fill in these categories. And I have to think about which category is which. I have to think why adjectives, and why objects and things, and it's interesting''} (F3).

\subsubsection{Challenge: Interface Complexity}

The block-based programming mode has more interface elements, which participants recognized as a learning curve. 
Additionally, even once familiar with the interface, there is still a time cost to creating longer or more complex programs where many components need to be edited.
R2 described: \textit{``So the disadvantage, of course, is like having to go through the whole process, which can be as long or short as necessary. It's the one that has the most, I would say, interaction cost in terms of just having to touch your device and manipulate the interface''} (R2).
Creating a long program like \texttt{`find TEXT on LICENSE PLATE on CAR, find COLOR on CAR'} would take more time than creating the program \texttt{`find PHONE',} simply because there are more parameters.

\subsection{Generating Programs from Questions}

Participants considered question mode to be the fastest way to get started making programs. However, the generated programs sometimes did not capture the correct parameters.

\subsubsection{Advantages: Fast and Approachable}

Many participants preferred the natural language mode because it was fast, and required the least cognitive effort. 
As R4 described:
\textit{``It's faster, overall like, you don’t have to break it down and add an adjective, or add this or that. You just have to have it categorize things correctly, which it seems like it mostly does''} (R4).
Even if the generated program was not exactly as intended, participants felt that they could use it as a starting point for editing (either with blocks or follow-up questions).
For example, F3 said, \textit{``I feel like I'm more in control of it. I can be just type in what I want, and then customize it from there. It just seems more straightforward to me to interact that way''} (F3).

\subsubsection{Challenge: Language Can Be Vague}

Participants felt that sometimes the generated programs did not capture their full intent.
R3 described how this could be due to ambiguities in natural language, comparing how they would phrase a program for finding dates versus one for finding bus routes: \textit{``I mean a date is a date. There’s no leniency. But if you’re asking for a bus route, do you mean the number of a route or the label of a route?''} (R3).
Because of this, some participants spent time thinking about exactly how to phrase a question to get it to detect what they wanted. R1 noted that they would rather edit the program directly than think about how to phrase questions: \textit{``[In block mode], I can be more specific and I can choose exactly what I want... Here, I don’t know, it’s still too difficult to phrase. I need too much brain power''} (R1).
Participants also sometimes forwent natural language, and simply dictated a statement in the program structure ("find blank on blank") to try to overcome this.

\subsection{Generating Programs from Examples}

Participant appreciated the potential of explore mode to give them a new awareness of visual features. But, in its current state, it was difficult for participants to know what features to select, and the generated programs sometimes contained unexpected conditions. 

\subsubsection{Advantages: Finding Unknown Features}

Participants saw the greatest strength of this mode as its potential to make them aware of new features and program possibilities.
\textit{``The summary gave information that I wasn't even aware of. So in that respect it was good, it was like `oh, buses are red, gotcha'''} (R3).
\textit{``For like the odd situation, you could use the explore option. And once you realize all of the details that are out there, you could say, well, this is worth categorizing, and I wanna be able to, you know, have a filter just based on stuff in the environment that I didn't know about. And I could actually see a real need for the the explore mode over the [question mode], because sometimes we don't know what we're working with in the field''} (F3).

Additionally, P3 pointed out that it was a useful way of testing the object detection would work on a specific item before manually writing a program for it. \textit{``For example, if there is some item, let's say medicine, or a bottle of milk or something that you consume every day. You don't know if the app would recognize that particular milk or not. So rather than try to create a program and try my luck, I can test it out directly. And if it detects, then you're cool, like, just create a filter for the next time, you know, and keep it saved, and then you can run it every time''} (P3).

\subsubsection{Challenge: Extrapolating Intent from a Single Feature}

In explore mode, participants needed to pick a feature of interest, and \sys generated a program based on finding that feature in the future.
Participants noted that it was hard to pick a target without knowing what the resulting filter would be, and without knowing the context of some visual features.
For example, when trying to generate a program that would read the route number on a bus, some participants debated between choosing the object `bus' or the number `73', both of which appeared as possible targets.

The program generation method we developed attempted to create filters to match specific visual content, though participant feedback revealed that one generation method is not suited to all tasks, and \sys may be including too many visual features. For example, R3 reasoned about why a generated filter included the adjective `red': \textit{``Maybe the bus in the image was red, but no it doesn't seem relevant to the route number. It probably did the right thing, it probably filtered it. Maybe all 73 buses are red. So it might have done the job that it was supposed to do, and not the job that I wanted it to do''} (R3).
F4 similarly noted that different features are relevant for different tasks. They generated a program by selecting `book' as the target, and the result was `find blue book on table': \textit{``You know you may be looking for. The colors, you're like, okay... When it comes to a book, that's not what I really need, because, like, when we look for our books, we look by title and all, of course''} (F4).
Considering these types of semantics when generating programs could improve the results.

\subsection{Comparing Creation Modes}

Participants generally appreciated all three programming modes were available in the app. An overview of how participants compared the modes is shown in Figure \ref{fig:ratings}.
Each mode required participants to think about their goal in a new way, and required different types of effort to turn it into an operable program.
Parameterizing a request in block mode, selecting a visual feature in explore mode, and phrasing requests in question mode each presented their own strengths and challenges.
Because of this, participants could see benefits of using different programming interfaces in different scenarios. As R4 put it: 
\textit{``It's all contextual, I think. So it depends on what you want to do. Like, we did 3 different examples. But I would use different methods based on what I knew about the environment. It just depends on what you're doing, you know, if you already have an image you're working with. You might go with that particular program, you know, you explore the image, and then that’s what you use. It just really depends on the situation''} (R4).

Participants also noted that the different modes could be helpful to people with different levels of technical expertise.
P3 said, \textit{``I think, also, like, to the user, they'll be scared, like, I have to do programming to use the app... But I mean, anyone pretty much can do it, and there are multiple ways, even if one method is gone. So there were two other methods to create''} (P3). 

\subsubsection{Structured vs. Unstructured Input}

Despite question mode seeming more approachable, block mode's structure provided participants a framework to work within. 
As P1 described: \textit{``I think one positive is that it gives you more of a predetermined list to choose from. So kinds of objects that you might be interested in, kinds of things on those objects that you might be looking for. One of the challenges with personalization is you can give someone too much choice to the point that they are overwhelmed and unsure of where to even start''} (P1).
This framework gave people an understanding of the limits of what they could create.

To balance these approaches, multiple participants imagined a hybrid block and natural language approach, where the block structure would still be present, but they could type in or dictate each program item instead of scrolling through menus. 
For example, F3 described, \textit{``Maybe if the display were slightly different, like `find any blank on any blank', and I could input the text there, that would make sense. But I'm trying to choose objects and adjectives and things like that. And it just seems a little cluttered''} (F3).

\subsubsection{Editing Generated Programs}

Participants generally appreciated that the two program generation modes (explore mode and question mode) displayed their results in the block interface, even participants who did not prefer that interface to start. For instance, F2 and F3 both preferred the generation interfaces, but agreed that it was easier to edit an existing program than to create one from scratch.
Participants also expressed that they would use this interface to refine the generated programs, as P4 said, \textit{``It's always good to have a backup there''} (P4).

\subsection{Benefits and Drawbacks of DIY-ing Assistive Technology}

Participants appreciated the deeper level of customization available when programming filters as compared to current assistive technology. 
As R2 described, it puts power into the hands of the user to decide what they needed:
\textit{``I think it all comes down to providing choice. Ultimately, what I like is that you're putting the information available, in the person's hands to choose... You know, just being able to empower people to be independent... What you've all created here is really neat because it's creating modularity to access the information. And I love that. I love that. And I wish more and more assistive technology companies thought about, how can we take these pieces of information and put it in the hands of the people that need it in a way that they can then modify it and change it and make it their own''} (R2).

On the other hand, R1 noted that having to put in the effort to create filters themselves could be considered a burden:
\textit{``People are not developers. Another developer, I suppose, knows how to program… I’m not a builder. There are tasks that are difficult for us to do, but what, I have to spend 5 hours to tinker with a program for what?''} (R1).
R1 expressed that they felt like \sys was a good option for developers wanting to quickly create things, but that it may not be ideally directed towards end-users.
Other participants who did not mind the idea of programming still mentioned that re-framing the functionality might make \sys feel more approachable.
Eventually, \sys could be considered as a platform for people to share programs that they have created, enabling a level of collaboration among end-users with different levels of expertise.

\section{Discussion and Future Work}

We found that \sys addresses unmet needs, empowering blind people to customize their experiences with AI.  
Here, we outline opportunities for building on \sys to further improve its utility.

\subsection{Raising the Ceiling of Creation Possibilities}

Throughout our user studies and analysis of existing data, we encountered many scenarios that could be addressed by \sys with the addition of new program operators. While \sys was implemented with `find' and `on' statements for simplicity and approachability, the addition of new operators could make \sys more expressive for DIY enthusiasts and power users.
For example, the addition of traditional logical operators such as \texttt{AND}, \texttt{OR}, and \texttt{NOT} would allow for a greater degree of specificity in programs. However, \texttt{AND} and \texttt{OR} are easily confused among end-users as their natural language counterparts can be ambiguous, so introducing these would need to be done with care.

New operators could also define additional ways for objects to interact with each other. Currently, `on' denotes objects whose bounding boxes are primarily overlapping. An operator like `nearby' could specify items that do not overlap, but are in proximity. Similarly, `following' could specifically find text content after a phrase, as in \texttt{find TEXT following "EXP:"} for more specifically finding an expiration date.
\sys as a system could be extended with these, but it would require new block interface designs, a challenge for approachability and accessibility.

Additionally, \sys could benefit from the inclusion of additional, specialized models for certain tasks.
For instance, a model that detects the make and model of a vehicle for locating a ride share, or text classification models that can filter out `brand names' or `flavors' for shopping scenarios. 
One notable object class missing in \sys is `digital display', for reading screens on thermostats, microwaves, or buses. We attempted to use YOLO-World to detect this class, but found that it was not accurate enough to be usable.
Although we currently use customized YOLO-World models on-device so the classes are pre-selected, YOLO-World was built for the ability to add new classes in real-time. In the future, when question mode extracts parameters from a request, the server could automatically generate new YOLO-World models to fill in gaps in the program as needed.

In the future, we imagine \sys being paired with other personalization approaches to create a deeper level of customization.
For instance, if combined with the capabilities of teachable object recognizers, users would not only be able to locate personal items, but to integrate them into programs as a basis for further filtering or automation.
Overall, we also see \sys as going beyond programming for visual information tasks. We believe our study reveals important findings about how blind end-users program, ideally leading to supporting more complex tools for people to DIY a range of assistive technologies outside of filtering programs.

\add{
Furthermore, when the creation ceiling in \sys is raised with things like additional operators and models, or if end-user programming approaches are eventually used to create different types of assistive technologies, this comes with a trade-off.
Balancing this complexity with ease of use is a critical concern for future accessibility work in this direction.
For instance, although \sys could eventually include a large library of models and classes to detect, these could be activated selectively or as-needed for different users or situations, making the system less overwhelming. Eventually, \sys could also make suggestions for how to create or improve programs based on how a person uses the system over time.
}

\vspace{-0.25pc}
\subsection{Automating Running Programs}

Because of the long-tail problem, some of our participants saw managing a library of programs as unwieldy.
For example, R1 said, \textit{``Would I have to program actions for all the objects in the world?''} (R1).
P1 expressed a similar sentiment: \textit{``When it comes to just the variety of information that anyone might be looking for, at any given point of time... I don't know. It just feels like there are so many permutations and combinations here. So many ways in which humans may want to query information that trying to build an even remotely comprehensive list of the more common categories of information seems like an endeavor that's really hard''} (P1).

Being able to automate when programs are run could remove some of this burden. For example, a \texttt{`find ADDRESS on PACKAGE'} program could be automatically started whenever a package enters the frame, on the assumption that the user is sorting mail. Or, like other mobile automations, programs could be tied to a location. For example, when a user arrives at a bus stop, the \texttt{`find NUMBER on BUS'} program could start.

\vspace{-0.25pc}
\subsection{Programming in the Age of VLMs}

Although large vision language models (VLMs) are becoming more powerful, they may not be a panacea, and making them truly beneficial requires deep integration with the needs of blind people. 
Despite the advantages of providing fully automated, subjective descriptions, they also add new challenges for blind users to acquire information. 
While this is still an area of active research, Massiceti et al. found the CLIP-based models were up to 15\% less accurate on images taken by blind people \cite{massiceti2023explaining}.
Additionally, as hallucinations seem to happen more often when describing complex scenes \cite{liu2024survey} or when being asked a follow-up question (the model appears to second guess itself), they could potentially arise more in accessibility contexts.

\add{Generally, this also calls back to the long lived direct manipulation vs interface agents debate \cite{manipulation-vs-agents}; although there is an effort cost to creating personalizations, there is also a cost when an intelligent system assumes someone’s needs and gets them wrong.}
Although they may appear at odds, we envision end-user programming as a supplement, not a replacement for large VLMs. We envision the two approaches complementing each other in the following ways:

\textbf{Reducing hallucinations by breaking down problems.} Programs can serve as a way to break down visual problems into smaller pieces, avoiding complex questions that might cause models to fail. Just as \sys crops each image frame to relevant object bounding boxes when running programs, a cropped version of an image could be passed to the VLM to query in a more constrained way. 
For example, programs could contain subjective adjectives like `clean' or `matching'. A VLM could be queried for these items in a constrained way, the answer could be parsed and fed back into the program. VLMs still are not good at reading large chunks of printed text or reasoning about complex scenes, but if the input and output were constrained then they may produce better results.

\textbf{Balancing ambiguity in language and improving explainability.} As discussed in our study findings, language is ambiguous, and programs can help articulate specific intents. Additionally, programs can serve as explicit step-by-step instructions of what a system is doing to come up with a given answer. This could help users better understand the limits of different tools, to better understand and predict why they fail.

\textbf{Making large VLMs `live' and creating reusable queries.} 
Current large VLMs take in a static image as input. Yet, as models become faster, running a query on a live camera feed will not be as simple as repeating the question on each frame.
Because programs specify what users want to hear and when, they could be used to convert natural language responses into real-time feedback.

\section{Conclusion}

We have presented \sys, an end-user programming tool for creating custom visual filtering programs. \sys implements a set of programming interfaces: block-based, natural language, and programming by example.
Through a user study of \sys conducted with 12 blind participants, we demonstrate the promise of end-user programming approaches for creating and customizing AI-based assistive technologies. We observed that users prefer different approaches depending on their experiences and the task, and also note areas where blind end-user programmers may face unique challenges while creating highly visual, camera-based technologies.
Overall, \sys is a step towards supporting blind people in creating personally meaningful assistive technologies.

\begin{acks}
We sincerely thank our participants for their time, and for sharing their expertise and experiences.
We also thank our reviewers for their time and feedback.
This research was supported in part by a Google Research Scholar Award.
This material is based upon work supported by the National Science Foundation Graduate Research Fellowship under Grant No. DGE-1841052.
Any opinion, findings, and conclusions or recommendations expressed in this material are those of the authors(s) and do not necessarily reflect the views of the National Science Foundation.
\end{acks}

\bibliographystyle{ACM-Reference-Format}
\bibliography{main}


\end{document}